\documentclass[preprint,showpacs,preprintnumbers,amsmath,amssymb]{revtex4}

\usepackage{graphicx}
\usepackage{dcolumn}
\usepackage{bm}

\begin{document}

\preprint{APS/123-QED}

\title{Superdeformed $\Lambda$ hypernuclei with antisymmetrized molecular dynamics}

\author{M. Isaka$^1$, K. Fukukawa$^2$, M. Kimura$^3$, E. Hiyama$^1$, H. Sagawa$^{1,4}$, and Y. Yamamoto$^1$}
\affiliation{$^1$RIKEN Nishina Center, RIKEN, Wako, Saitama 351-0198,Japan}
\affiliation{$^2$Instituto Nazionale di Fisica Nucleare, Sezione di Catania, Via Santa Sofia 64, I-95123 Catania, Italy}
\affiliation{$^3$Department of Physics, Hokkaido University, Sapporo 060-0810,Japan}
\affiliation{$^4$Center for Mathematics and Physics, University of Aizu, Aizu-Wakamatsu, Fukushima 965-8560,Japan}

%

\date{\today}

\begin{abstract}
The response to the addition of a $\Lambda$ hyperon is investigated for the deformed states such as superdeformation in $^{41}_\Lambda$Ca, $^{46}_\Lambda $Sc and $^{48}_\Lambda$Sc. In the present study, we use the antisymmetrized molecular dynamics (AMD) model. It is pointed out that many kinds of deformed bands appear in $^{45}$Sc and $^{47}$Sc. Especially, it is found that there exists superdeformed states in $^{45}$Sc. By the addition of a $\Lambda$ particle to $^{40}$Ca, $^{45}$Sc and $^{47}$Sc, it is predicted, for the first time, that the superdeformed states exist in the hypernuclei $^{41}_\Lambda$Ca and $^{46}_\Lambda$Sc. The manifestation of the dependence of the $\Lambda$-separation energy on nuclear deformation such as spherical, normal deformation and superdeformation is shown in the energy spectra of $^{41}_\Lambda$Ca, $^{46}_\Lambda $Sc and $^{48}_\Lambda$Sc hypernuclei.
\end{abstract}

\pacs{Valid PACS appear here}
\maketitle

\section{Introduction}

One of the main purpose to study the structure of hypernuclei is to investigate nuclear responses when a $\Lambda$ particle is added to the nucleus. Since a $\Lambda$ hyperon is free from the nuclear Pauli principle, we have many interesting phenomena in hypernuclei. 
For example, in light $p$-shell hypernuclei with $A \leq 10$, it was pointed out that the $\Lambda$ participation gives rise to more bound states and the appreciable contraction of the system (this stabilization is called the 'gluelike' role of the $\Lambda$ particle) by cluster model \cite{PTP70.189(1983),PTPS81.42(1985),PRC59.2351(1999)}. 
In heavier $p$-shell ($A \geq 10$) and $sd$-shell $\Lambda$ hypernuclei, there is a huge variety of nuclear structure, because they have both $\alpha$-clustering and (compacting and/or deformed) shell-model like structures. 
For example, in $^{13}_\Lambda$C, we found the dynamical contraction of the clustering states as much as 30 \% by addition of a $\Lambda$ hyperon, while in the compacting shell-model like state, there was almost no dynamical shrinkage \cite{PTP97.881(1997),PRL85.270(2000)}. 

It is also interesting to investigate the difference of the $\Lambda$-separation energy ($B_\Lambda$) between the cluster and shell-model like states.
For example, in $^{13}_\Lambda$C, it is found that the $B_\Lambda$ is smaller in the clustering state than in the compacting shell-model like states by more than 3 MeV \cite{PTP97.881(1997)}. 
In $^{20}_\Lambda$Ne, it is predicted that the $B_\Lambda$ is larger in the shell-model like ground state than in the negative-parity cluster state \cite{PRC83.044323(2011)}. 
However, this result is different from the cluster model calculation \cite{PTP78.1317(1987)}. 
Therefore it is important to reveal differences of $B_\Lambda$ depending on the structures.

Let us consider the heavier nuclei with $20 \le A \le 40$. 
It is of interest to investigate nuclear response when a $\Lambda$ particle is injected into such heavier nuclei. In these nuclei, there are various interesting phenomena such as the coexistence of spherical, normal deformed (ND) and clustering states in the low-lying energy regions. Therefore it is expected that a $\Lambda$ particle generates the hypernuclear states with various structures and modifies them. For example, it is pointed out that the shrinking effect of $\alpha$-clustering states will appear by adding a $\Lambda$ particle \cite{ PTP71.985(1984),PRC83.054304(2011)}. It is predicted that the shape of $^{29}_\Lambda$Si is changed to be spherical by a $\Lambda$ particle, whereas the corresponding core nucleus $^{28}$Si is oblately deformed \cite{PRC78.054311(2008)}. 

It is interesting to study structure of Sc isotopes, because of the coexistence of spherical and normal deformed states. 
Many authors studied the structure of odd Sc isotopes, in which the positive-parity and negative-parity bands coexist near the ground state.   
This is inconsistent with the naive shell-model picture which suggests only the negative-parity states near the ground state. 
For example, $^{45}$Sc has the degenerated positive-parity state  $3/2^+$ near the ground state ($7/2^-$)  with $\Delta E = 12 $keV. 
In $^{47}$Sc, the positive-parity state $3/2^+$ lies at only 0.77 MeV above the ground state $7/2^-$. 
The occurrence of the low-lying positive-parity states can be explained as a proton 2-particle 1-hole ($2p$-$1h$) state in the Nilsson model \cite{ Phys.Scr8.135(1973),NuovoCim.26A.25(1975),NPA262.317(1976),Act.Phys.Pol.B16.847(1985)}. 
In other words, it could be said  that the low-lying positive-parity states are deformed, while the ground states $7/2^-$ are almost spherical. 
By adding a $\Lambda$ hyperon to them, it is expected that the low-lying excitation spectra will be modified in Sc $\Lambda$ hypernuclei, because the $B_\Lambda$ will be different depending on the deformation. 
Especially, in the case of the hyperon in $s$-wave, it is predicted that the $\Lambda$-separation energy $B_\Lambda$ becomes larger for the spherical state than the deformed state \cite{PRC78.054311(2008),PRC83.014301(2011),PRC84.014328(2011),PRC83.044323(2011),PTP128.105(2012)}. Therefore, the addition of a $\Lambda$ particle will affect the low-lying spectra of Sc isotopes, and hence, we expect that Sc isotopes are suited to verify the dependence of $B_\Lambda$ on nuclear deformation, which is one of the purposes of this study.

Another interesting issue in $\Lambda$ hypernuclei with $A \sim 40$ is to study the low-lying states with very large deformation, $i.e.$ superdeformed (SD) states. 
It is considered that the SD state is related with the appearance of large shell gaps at a $2:1$ axis ratio of the nuclear deformation for specific particle numbers.  
Recently, many SD bands are investigated and identified in nuclei with $A \sim 40$ \cite{PRL85.2693(2000),PRL87.222501(2001),PRC61.064314(2000)}.
For example, in $^{40}$Ca, a SD band is confirmed by the observation of the $K^\pi = 0^+$ band built on the $0^+_3$ state lying at 5.21 MeV , while its ground state is spherical due to the double closed-shell structure \cite{PhysRevC3.219(1971),PRL87.222501(2001)}. 
It is expected that in some Sc isotopes, the SD states will appear in the low-lying regions, which could be accessed by 
 the $^{24}$Mg + $^{24}$Mg reaction fusion-evaporation experiment at GANIL \cite{Ideguchi}.

Second purpose of the present study is difference of $B_\Lambda$ which indicates difference of deformations.
To investigate it, we perform the antisymmetrized molecular dynamics (AMD) calculations for $^{40}$Ca, $^{45}$Sc and $^{47}$Sc and corresponding $\Lambda$ hypernuclei. The AMD model was applied successfully to several normal nuclei with $A \sim 40$ such as $^{40}$Ca \cite{PRC72.064322(2005),PRC76.044317(2007)} to reveal the structure of the SD states without assumptions on nuclear deformations and clustering.
In this study, we apply the same AMD framework as done in Ref. \cite{PRC76.044317(2007)} so as to predict the existence of the ND and SD bands in $^{45}$Sc and $^{47}$Sc. 
In hypernuclear study, an extended version of the AMD model for hypernuclei (HyperAMD) is applied to investigate the hypernuclear SD states.   It was known that  the HyperAMD 
succeeded to describe the low-lying structure of $p$ and $sd$ shell $\Lambda$ hypernuclei  combined with the generator coordinate method (GCM) \cite{PRC83.054304(2011)}. Thus  the HyperAMD is 
confirmed as a powerful tool to study the structure of the $\Lambda$ hypernuclei with $A \sim 40$ having various deformations.
It is very important to use a reliable $\Lambda N$ interaction for calculations of $B_\Lambda$ values. We adopt here the $\Lambda N$ G-matrix interaction derived from the extended soft-core (ESC) model \cite{PTPS185.14(2010)}, which reproduces $B_\Lambda$ values of $\Lambda$ hypernuclei systematically \cite{PTPS185.72(2010)}. 
The present study is very much motivated by an experimental project at JLab which is planned to produce from $s$-shell with $A \sim 4$ up to $A \sim 208$ $\Lambda$ hypernuclei by the $(e, e' K^+)$ reaction. 
There is high expectation  to observe the SD states in $^{46}_\Lambda$Sc and $^{48}_\Lambda$Sc by the $^{46}$Ti ($^{48}$Ti) $(e, e' K^+)$ at JLab.

This paper is organized as follows. In the next section, we explain the theoretical framework of HyperAMD. 
In Section \ref{sec3}, the changes of the $\Lambda$ single particle energy as a function of quadrupole deformation are presented. The difference of the $\Lambda$-separation energy in $^{41}_\Lambda$Ca among the ground ND and SD states is also discussed in this section.
In Section \ref{sec4}, the existence of the hypernuclear SD states and the changes of the excitation spectra are discussed for $^{46}_\Lambda$Sc and $^{48}_\Lambda$Sc. The final section summarizes this work.

\section{Framework}

In this study, we apply the HyperAMD combined with the generator coordinate method (GCM) \cite{PRC83.054304(2011)} to $^{41}_\Lambda$Ca, $^{46}_\Lambda$Sc and $^{48}_\Lambda$Sc hypernuclei. 

\subsection{Hamiltonian and variational wave function}

The Hamiltonian used in this study is given as,
\begin{eqnarray}
\hat{H} &=& \hat{H}_{N} + \hat{H}_{\Lambda} - \hat{T}_g, \\
\label{H_N}
\hat{H}_{N} &=& \hat{T}_{N} + \hat{V}_{NN} + \hat{V}_{Coul},\\
\label{H_L}
\hat{H}_{\Lambda} &=& \hat{T}_{\Lambda} + \hat{V}_{\Lambda N}.
\end{eqnarray}
Here, $\hat{T}_{N}$, $\hat{T}_{\Lambda}$ and $\hat{T}_g$ are the kinetic energies of nucleons, a $\Lambda$ hyperon and the center-of-mass motion, respectively.
We use the Gogny D1S interaction \cite{PRC21.1586(1980)} as an effective nucleon-nucleon interaction $\hat{V}_{NN}$. 
In $^{45}$Sc and $^{47}$Sc, we tune the spin-orbit interaction of Gogny D1S to reproduce the observed excitation energy of the $3/2^+_1$ state of $^{45}$Sc.
The Coulomb interaction $\hat{V}_{Coul}$ is approximated by the sum of seven Gaussians. We use the $G$-matrix interaction as a $\Lambda$N effective interaction $\hat{V}_{\Lambda N}$. 

The intrinsic wave function of a single $\Lambda$ hypernucleus composed of a core nucleus with mass $A$ and a $\Lambda$ hyperon is described by the parity-projected wave function, $\Psi^\pi = \hat{P}^\pi \Psi_{int}$, where $\hat{P}^\pi$ is the parity projector and $\Psi_{int}$ is the intrinsic wave function given as, 
\begin{eqnarray}
\Psi_{int} &=& \Psi_N \otimes \varphi_\Lambda,\quad
\Psi_N = \frac{1}{\sqrt{A!}}\det \left\{ \phi_{i} \left( r_j \right) \right\},\\
\phi_{i} &=& \prod_{\sigma=x,y,z} \biggl(\frac{2\nu_\sigma}{\pi}\biggr)^{\frac{1}{4}}
 \exp \biggl\{-\nu_\sigma \bigl(r - Z_{i} \bigr)_\sigma^2 \biggr\} \chi_{i} \eta_{i} ,\\
\varphi_\Lambda &=& \sum_{m=1}^M c_m \chi_m \prod_{\sigma=x,y,z} \biggl(\frac{2\nu_\sigma}{\pi}\biggr)^{\frac{1}{4}}
 e^{-\nu_\sigma \bigl(r - z_m \bigr)_\sigma^2},
\end{eqnarray}
\begin{eqnarray}
\chi_{i} &=& \alpha_i \chi_\uparrow + \beta_i \chi_\downarrow,\quad \chi_m = a_m \chi_\uparrow + b_m \chi_\downarrow, \\
\eta_{i} &=& {\rm proton}\ {\rm or}\ {\rm neutron},
\end{eqnarray}
where $\phi_{i}$ is {\it i}th nucleon single-particle wave packet consisting of spatial, spin $\chi_{i}$ and isospin $\eta _{i}$ parts. 
The variational parameters are the centroids of Gaussian $\bm{Z}_i$ and $\bm{z}_m$, width parameters $\nu_\sigma$, spin directions $\alpha_i$, $\beta_i$, $\alpha_m$ and $\beta_m$, and coefficients $c_m$. We approximately remove the spurious center-of-mass kinetic energy in the same way as Ref. \cite{PRC83.044323(2011)}.

\subsection{Variation with constraint on nuclear quadrupole deformation $\beta$}

The energy variation is performed under the constraint on nuclear matter quadrupole deformation parameter $\beta$ in the same way as our previous works \cite{PRC83.044323(2011),PRC83.054304(2011)}. 
By the frictional cooling method, the variational parameters in $\Psi^\pi$ for the $\Lambda$ hypernucleus and $\Psi^\pi_N$ for the core nucleus are determined for each $\beta$, and the resulting wave functions are denoted as $\Psi^\pi (\beta)$ and $\Psi^\pi_N (\beta)$, respectively. 
It is noted that the nuclear quadrupole deformation parameter $\gamma$ is optimized through the energy variation for each $\beta$.
By using the wave functions$\Psi^\pi (\beta)$ and $\Psi^\pi_N (\beta)$, we calculate the parity-projected intrinsic energies defined as,
\begin{eqnarray}
\label{int-H}
E^\pi_{hyp} (\beta) &=& \langle \Psi^{\pi}(\beta) | \hat{H} | \Psi^{\pi}(\beta) \rangle / \langle \Psi^{\pi}(\beta) |\Psi^{\pi}(\beta) \rangle,\\
\label{int-C}
E^\pi_{cor} (\beta) &=& \langle \Psi^{\pi}_N(\beta) | \hat{H} | \Psi^{\pi}_N(\beta) \rangle / \langle \Psi^{\pi}_N(\beta) |\Psi^{\pi}_N(\beta) \rangle,
\end{eqnarray}
for the $\Lambda$ hypernucleus and the core nucleus, respectively.

It is found that the $\Lambda$ hyperon dominantly occupies an $s$-wave in the hypernuclear states, because no constraint is imposed on the $\Lambda$ single particle wave function in the present calculation. Combined with the parity projection, we obtained two kinds of configurations in which the $\Lambda$ hyperon couples to the positive- and negative-parity states of the core. Here these two are denoted $\Psi^+_N \otimes \Lambda$ and $\Psi^-_N \otimes \Lambda$, respectively, in the following discussions.

\subsection{Angular momentum projection and GCM}

After the variational calculation, we project out an eigenstate of the total angular momentum from the hypernuclear states, 
\begin{eqnarray}
 \label{AngProj}
 \Psi^{J\pi}_{MK}(\beta_i) &=& \frac{2J+1}{8\pi^2} \int d\Omega D^{J*}_{MK}(\Omega) \hat{R}(\Omega) \Psi^{\pi} (\beta_i).
\end{eqnarray}
The integrals are performed numerically over three Eular angles $\Omega$.

The wave functions $\Psi^{J \pi}_{MK} (\beta_i)$ which have the same parity and angular momentum but have different $K$ and nuclear quadrupole deformation $\beta_i$ are superposed (GCM). 
Then the wave function of the system is written as
\begin{eqnarray}
\label{GCM_wf}
\Psi^{J\pi}_{\alpha} &=& c_\alpha \Psi^{J \pi}_{MK}(\beta) + c'_\alpha \Psi^{J \pi}_{MK'}(\beta') + \cdots ,
\end{eqnarray}
where the quantum numbers except for the total angular momentum and the parity are represented by $\alpha$. The coefficients $c_\alpha$, $c'_\alpha$, $\cdots$ are determined by the Hill-Wheeler equation.

%

\subsection{Analysis of wave function}

It is convenient to define the $\Lambda$ single particle energy $\epsilon_\Lambda$ as,
\begin{eqnarray}
\epsilon_\Lambda (\beta) = \langle \Psi^\pi (\beta) | \hat{H_\Lambda} | \Psi^\pi (\beta) \rangle / \langle \Psi^\pi (\beta) | \Psi^\pi (\beta) \rangle,
\label{LMDsp}
\end{eqnarray}
to discuss how the $\Lambda$ binding energy changes for increasing the nuclear deformation.

We also calculate the energy gain $B_\Lambda$ of each state $J^\pi$ in a hypernucleus from the corresponding state of the core nucleus as follows. 
\begin{eqnarray}
B_\Lambda = E(^{A}{\rm Z} (j^\pi) ) - E(^{A+1}_\Lambda{\rm Z} (J^\pi)).
\label{bLmd}
\end{eqnarray}
Here, $E(^{A}{\rm Z}) (j^\pi)$ and $E(^{A+1}_\Lambda{\rm Z})(J^\pi)$ respectively represent the total energies of the $j^\pi$ state of the core nuclei and the corresponding $J^\pi$ state of the hypernuclei.
$B_\Lambda$ shows the $\Lambda$-separation energy for each state of hypernuclei.

To analyze the nucleon configurations for the wave functions on energy curves, we calculate the number of the deformed harmonic oscillator quanta for protons ($N^p$) and neutrons ($N^n$) in the same way as Refs. \cite{PRC75.034312(2007),PRC83.044304(2011)}. The number operators are defined as, 
\begin{eqnarray}
\hat{N}^{p (n)} &=& \sum_i \sum_{\sigma = x,y,z }\left( \frac{p^2_{\sigma i}}{4\hbar^2 \nu_\sigma} + \nu_\sigma r^2_{\sigma i} -\frac{3}{2} \right), 
\label{NumHOQ}
\end{eqnarray}
where summation over $i$ runs all protons (neutrons) \cite{PRC75.034312(2007)}. 
Here the expectation values of $\hat{N}^\xi$, $N^\xi = \langle \Psi^\pi_N |\hat{N}^\xi| \Psi^\pi_N \rangle/\langle \Psi^\pi_N | \Psi^\pi_N \rangle$ are calculated for the parity projected states of the core nuclei.

We introduce the overlap between the GCM wave function $\Psi^{J\pi}_\alpha$ and $\Psi^{J \pi}_{MK} (\beta)$, 
\begin{eqnarray}
O^{J\pi}_{MK\alpha} ( \beta ) = | \langle \Psi^{J \pi}_{MK} ( \beta ) | \Psi^{J \pi}_\alpha \rangle |^2.
\label{Overlap}
\end{eqnarray}
Since the GCM overlap $ O^{J\pi}_{MK \alpha} ( \beta )$ shows the contributions from $\Psi^{J \pi}_{MK} ( \beta )$ to each state $J^\pi$, it is useful to estimate the nuclear quadrupole deformation $\beta$ of each state. 
Namely, we regard $\beta$ corresponding to the maximum GCM overlap as the nuclear deformation of each state.
We call the deformed states with $\beta < 0.5$ ND, and those with $\beta \ge 0.5$ SD. Detail discussions about ND and SD will be performed in the next Section.

\subsection{$\Lambda$N effective interaction}

It is reasonable to use an effective $\Lambda N$ interaction
in our model where short-range and tensor correlations 
are not included. The G-matrix theory makes it possible to
derive a realistic effective interaction in a model space, 
starting from a free-space interaction.
Baryon-baryon interaction models of the $SU(3)$ flavor-octet baryons
have been developed with use of $YN$ scattering data and also
hypernuclear information. 
One of them is the extended-soft-core (ESC) model~\cite{PTPS185.14(2010)} 
by the Nijmegen group, and the latest version is called ESC08c~\cite{ESC08c}.
Almost all hypernuclear data are reproduced consistently by ESC08c.
In this study, we use the $YN$ G-matrix interaction (YNG) derived
from ESC08c. The YNG is obtained in nuclear matter and represented in
a three-range Gaussian form:
\begin{eqnarray}
\label{YNG}
G(r;k_F)= \sum_{i=1}^3\, (a_i+b_i k_F+c_i k_F^2)\, \exp(-r^2/\beta_i^2) \ .
\end{eqnarray}
Here, an interaction strength in each $\Lambda N$ state depends on 
a nuclear Fermi momentum $k_F$, reflecting the density-dependence of
the G-matrix in nuclear medium. It should be noted that
the $\Lambda N$-$\Sigma N$ coupling term included in ESC08c is 
renormalized into the $\Lambda N$-$\Lambda N$ part of the G-matrix 
interaction, giving rise to the important part of the density dependence.
Values of parameters $(a_i,b_i,c_i)$ and $\beta_i$ are given in Appendix.

In the case of using the YNG interaction, decisively important is
how to treat its $k_F$-dependence in a structure calculation.
In Ref.\cite{PTPS185.72(2010)}, the averaged-density approximation (ADA) was demonstrated
to be very reliable to reproduce spectra of $\Lambda$ hypernuclei with
YNG interactions, where $k_F$ values in YNG were obtained from
expected values of nuclear density distributions by $\Lambda$
wave functions.

According to the ADA treatment, the $k_F$ value is calculated for each configuration of $^{41}_\Lambda$Ca, $^{46}_\Lambda$Sc and $^{48}_\Lambda$Sc by using the wave function $\Psi^\pi (\beta_0)$, where $\beta_0$ corresponds to the maximum of the GCM overlap. 
It is found that the obtained $k_F$ values are the same for the $\Psi^-_N \otimes \Lambda$ and $\Psi^+_N \otimes \Lambda$ states of each Sc hypernuclei. Finally, we take $k_F = 1.26$ fm$^{-1}$ for $^{41}_\Lambda$Ca, $k_F = 1.29$ fm$^{-1}$ for $^{46}_\Lambda$Sc, and $k_F = 1.30$ fm$^{-1}$ for $^{48}_\Lambda$Sc, respectively, in the GCM calculations. 

\section{Difference of $B_\Lambda$ among the spherical, ND and SD states in $^{41}_\Lambda {\rm Ca}$}
\label{sec3}

\begin{figure*}
  \begin{center}
    \includegraphics[keepaspectratio=true,width=172mm]{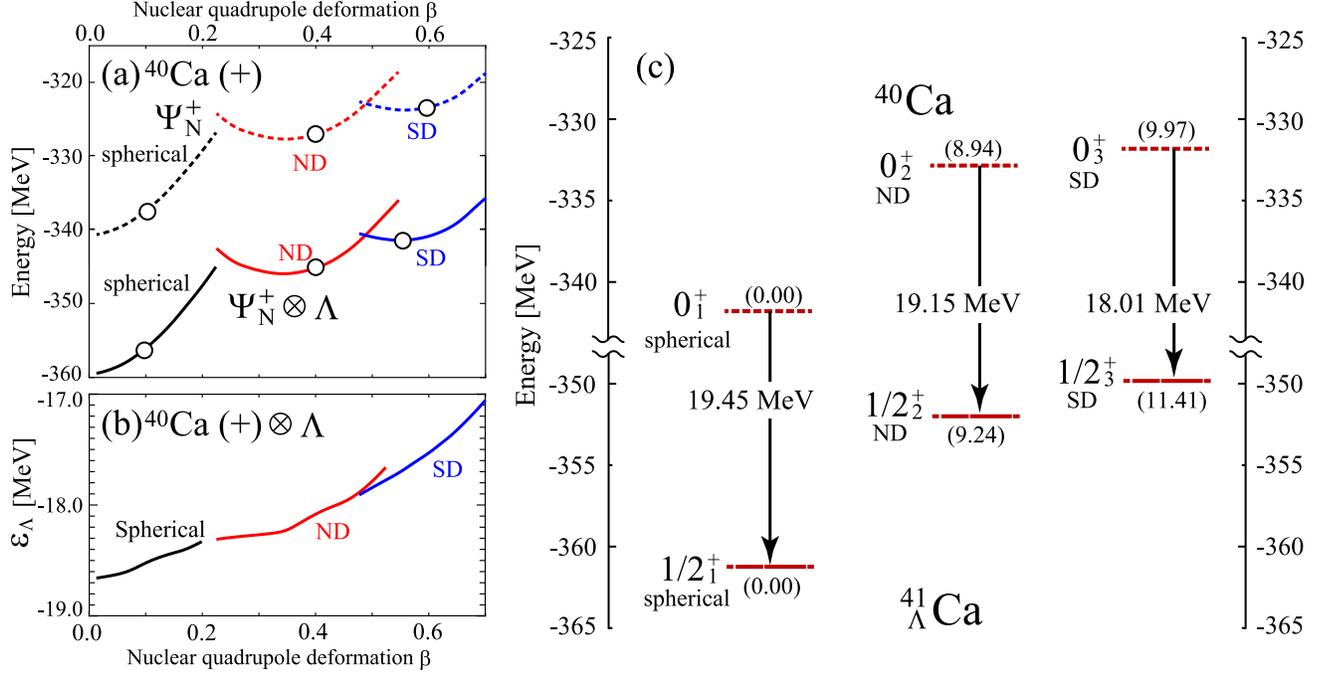}
  \end{center}
  \caption{(Color online) (a) Energy curves as a function of the nuclear quadrupole deformation $\beta$ for $^{40}$Ca (dotted) and $^{41}_\Lambda$Ca (solid). Deformation parameter $\gamma$ is optimized through the energy variation for each $\beta$. Open circles show the deformations $\beta$ of the $0^+$ and $1/2^+$ states calculated by the GCM corresponding to the energy curves. (b) $\Lambda$ single particle energy defined by Eq. (\ref{LMDsp}) for $^{41}_\Lambda$Ca. (c) Calculated energies of the $0^+$ states in $^{40}$Ca (dotted) and $1/2^+$ states in $^{41}_\Lambda$Ca (solid). The values in parenthesis are the calculated excitation energies from the ground states of $^{40}$Ca and $^{41}_\Lambda$Ca in MeV, respectively.}
  \label{fig:fig1.eps}
\end{figure*}

\begin{table*}
  \caption{Calculated excitation energy $E_x$ in MeV, matter quadrupole deformation $\beta$ and $\gamma$ [deg], and RMS radii [fm]. The $B_\Lambda$ defined by Eq. (\ref{bLmd})is also listed in unit of MeV for $^{41}_\Lambda$Ca.}
  \label{Tab:table1}
  \begin{ruledtabular}
  \begin{tabular}{cccccccccc}
                     &   $J^\pi$ & $E_x$[MeV] &$\beta$& $\gamma$[deg] & $r_{{\rm RMS}}$[fm] & $B_\Lambda$ \\
  \hline
   $^{41}_\Lambda$Ca & $1/2^+_1$ &  0.00 &  0.10 &  0 & 3.38 & 19.45 \\
                     & $1/2^+_2$ &  9.24 &  0.40 & 27 & 3.47 & 19.15 \\
                     & $1/2^+_3$ & 11.41 &  0.55 & 13 & 3.58 & 18.01 \\
  \\
   $^{40}$Ca         &  $0^+_1$  &  0.00 &  0.12 & 12 & 3.39 & \\
                     &  $0^+_2$  &  8.94 &  0.40 & 28 & 3.50 & \\
                     &  $0^+_3$  &  9.97 &  0.60 & 17 & 3.63 & \\
  \end{tabular}
  \end{ruledtabular}
\end{table*}

First, we show the results of $^{40}$Ca and $^{41}_\Lambda$Ca. 
Because there exist the ND and SD states in $^{40}$Ca, it is of interest to discuss the difference of the $\Lambda$-separation energy in $^{41}_\Lambda$Ca.
Here we briefly explain the properties of the deformed states of $^{40}$Ca.
It is observed that in $^{40}$Ca there exist the ND and SD bands built on the $0^+_2$ (at 3.35 MeV) and $0^+_3$ (5.21 MeV) states, respectively \cite{PhysRevC3.219(1971),PRL87.222501(2001)}. The SD band is considered to have the $[(sd)^{-4} (pf)^4]_\pi [(sd)^{-4} (pf)^4]_\nu$ configuration, while the configuration of the ND band is $[(sd)^{-2} (pf)^2]_\pi [(sd)^{-2} (pf)^2]_\nu$ \cite{NPA93.110(1967),NPA123.241(1969),PRL87.222501(2001)}. 
Our results for $^{40}$Ca shown in Fig. \ref{fig:fig1.eps} are identical to those reported in Ref. \cite{PRC76.044317(2007)}, since the same theoretical model and effective interaction are applied. 
The energy curve of $^{40}$Ca as a function of the deformation parameter $\beta$ (dotted line in Fig. \ref{fig:fig1.eps} (a)) shows the spherical energy minimum having the $sd$ closed-shell configuration, which corresponds to the ground state. It also shows two well-determined local energy minima at $\beta = 0.35$ and $\beta = 0.55$ which correspond to the ND and SD states of $^{40}$Ca mentioned above, respectively. 

Furthermore, in Fig. \ref{fig:fig1.eps} (a), we show the energy curve of $^{41}_\Lambda$Ca, which is similar in that of $^{40}$Ca. 
In Fig. \ref{fig:fig1.eps} (b), the $\Lambda$ single-particle energies with respect to nuclear quadrupole deformation $\beta$ are illustrated. We see that the $\Lambda$ single-particle energy $\epsilon_\Lambda (\beta)$ varies depending on $\beta$, $i.e.$ the $\epsilon_\Lambda$ has the minimum at $\beta = 0$, and becomes shallower as $\beta$ increases. This means that a $\Lambda$ hyperon is most deeply bound to the spherical shape core nucleus, which is consistent with the previous works for $p$-and $sd$-shell hypernuclei such as $^{13}_\Lambda$C and $^{21}_\Lambda$Ne \cite{PRC83.044323(2011),PRC83.014301(2011)}.

The difference of $\epsilon_\Lambda$ will lead to the difference of $B_\Lambda$ and affect the excitation spectra of $\Lambda$ hypernuclei. 
To investigate it, we perform the GCM calculation for $^{40}$Ca and $^{41}_\Lambda$Ca by using the wave functions on the energy curves.
Figure \ref{fig:fig1.eps}(c) and Tab. \ref{Tab:table1} show the calculated energies of three $1/2^+$ states of $^{41}_\Lambda$Ca, in which a $\Lambda$ in $s$-orbit is coupled to the ground, ND and SD $0^+$ states of $^{40}$Ca. 
We also calculate the $B_\Lambda$, defined by Eq. (\ref{bLmd}), for each state. It is found that the $\Lambda$-separation energy for the ground state $1/2^+_1$ is $B_\Lambda = 19.45$ MeV which is consistent with the empirical formula of $B_\Lambda$ \cite{ANP8.1(1975)}.
The $B_\Lambda$ is the largest for the spherical ground state, 19.45 MeV, and smallest for the SD $1/2^+_3$ state, 18.01 MeV. We see the energy difference about 1.4 MeV between the spherical and SD states.

Let us see the changes of radii and deformation by the $\Lambda$ addition. 
As shown in Tab. \ref{Tab:table1}, we do not have any difference in radii and deformations for the ground and ND states. On the other hand, in the SD state, it is seen that shrinkage of the radius by about 1.4 \% and the reduction of nuclear quadrupole deformation.

\section{Scandium hypernuclei}
\label{sec4}

\begin{figure}
  \begin{center}
    \includegraphics[keepaspectratio=true,width=86mm]{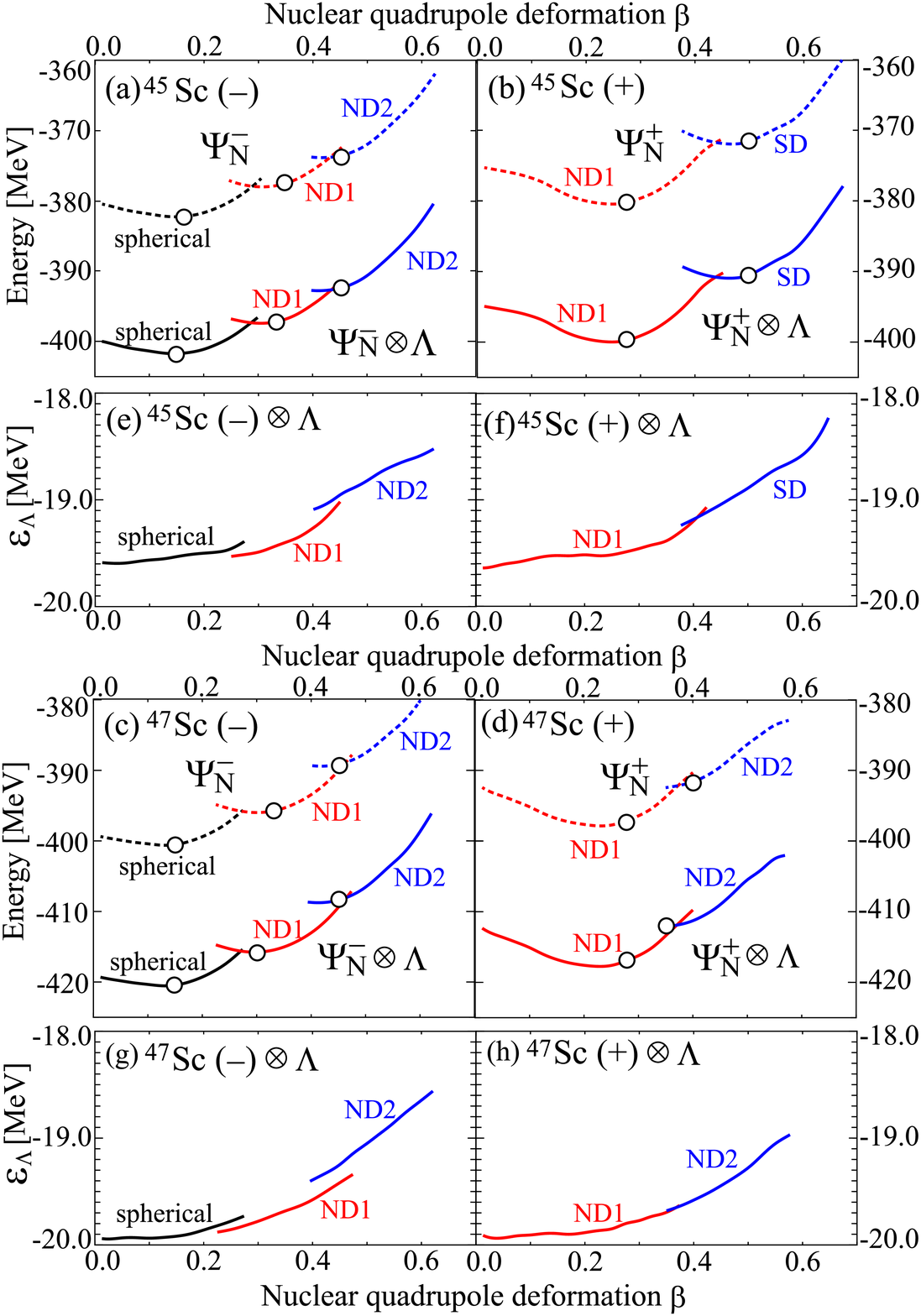}
  \end{center}
  \caption{(Color online) Same as Fig. \ref{fig:fig1.eps} (a) and (b), for $^{45}$Sc and $^{47}$Sc and the corresponding hypernuclei. Open circles show the deformation $\beta$ of the corresponding states obtained after the GCM calculation.}
  \label{fig:fig2.eps}
\end{figure}

\begin{figure}
  \begin{center}
    \includegraphics[keepaspectratio=true,width=86mm]{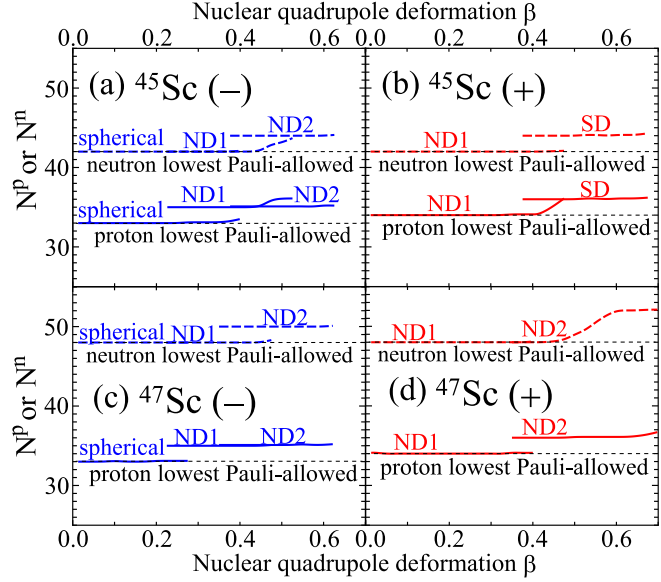}
  \end{center}
  \caption{(Color online) Numbers of the deformed harmonic oscillator quanta $N^p$ (solid) and $N^n$ (dashed), defined by Eq. (\ref{NumHOQ}),  calculated from the wave functions on the energy curves shown in Figure \ref{fig:fig2.eps}. (a) and (b) ((c) and (d)) correspond to the negative- and positive-parity states of $^{45}$Sc ($^{47}$Sc), respectively.}
  \label{fig:fig3.eps}
\end{figure}

\begin{figure}
  \begin{center}
    \includegraphics[keepaspectratio=true,width=86mm]{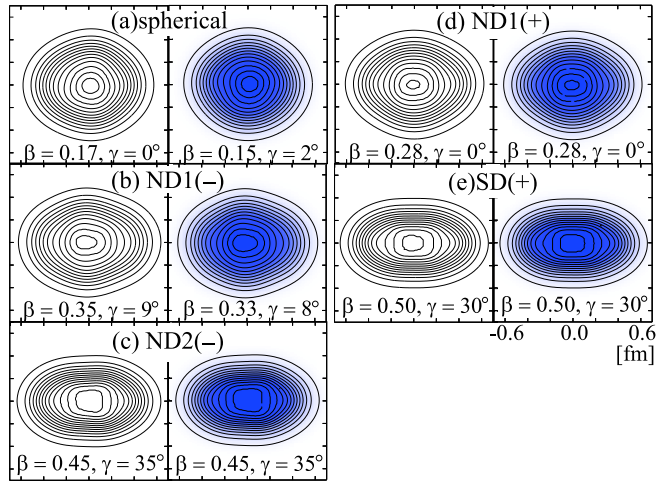}
  \end{center}
  \caption{(Color online) (a)-(c): intrinsic density distributions of the negative-parity spherical, ND1 and ND2 states in $^{45}$Sc and $^{46}_\Lambda$Sc, respectively. In each panels, contour lines show the nuclear density, while the color plots present the distribution of $\Lambda$. Values $\beta$ and $\gamma$ are deformations of each state determined by the GCM overlap. (d)-(e): same as the panels (a)-(c) for the positive-parity ND1 and SD states of $^{45}$Sc and $^{46}_\Lambda$Sc.}
  \label{fig:dens.eps}
\end{figure}

\begin{figure*}
  \begin{center}
    \includegraphics[keepaspectratio=true,width=172mm]{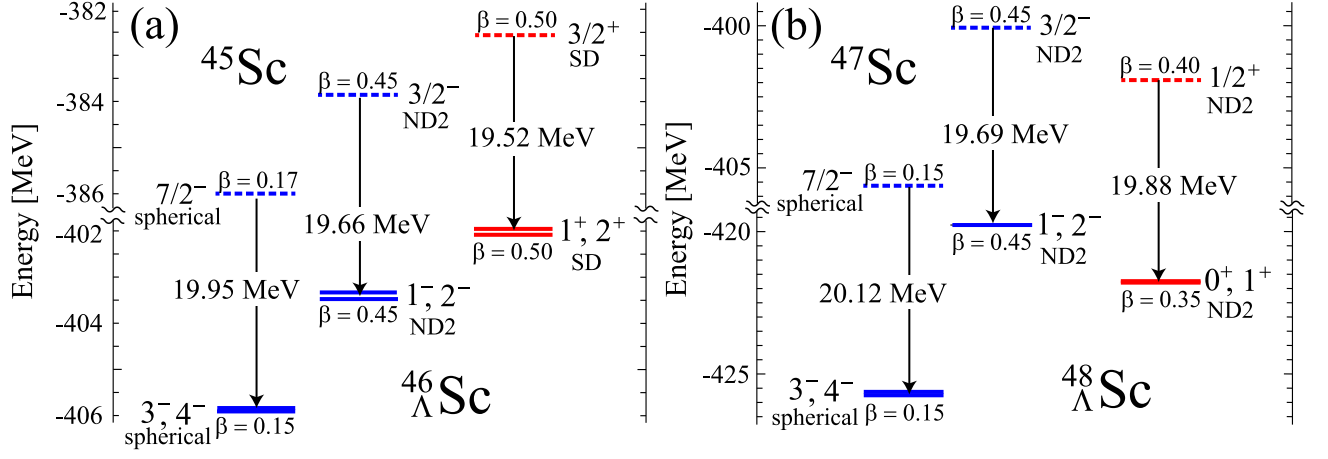}
  \end{center}
  \caption{(Color online) (a) Calculated energies of the ground and deformed states corresponding to the spherical, ND2 and SD minima of $^{45}$Sc and $^{46}_\Lambda$Sc shown in Fig. \ref{fig:fig2.eps}(a)-(b). (b) Same as (a) for the spherical and ND2 minima of $^{47}$Sc and $^{48}_\Lambda$Sc shown in Fig. \ref{fig:fig2.eps}(c)-(d). }
  \label{fig:fig4.eps}
\end{figure*}

\begin{table*}
  \caption{Same as Table \ref{Tab:table1} for $^{46}_\Lambda$Sc and $^{48}_\Lambda$Sc, and corresponding nuclei.}
  \label{Tab:table5}
  \begin{ruledtabular}
  \begin{tabular}{cccccccc}
 &  & $J^\pi$ & $E_x$[MeV] & $\beta$ & $\gamma$[deg] & $r_{{\rm RMS}}$[fm] & $B_\Lambda$[MeV] \\
 \hline
$^{46}_\Lambda$Sc 
 & spherical & $4^-_1$ & 0.00 & 0.15 & 2 & 3.51 & 19.95 \\
 & ND2 & $2^-_4$ & 2.47 & 0.45 & 35 & 3.61 & 19.66 \\
 & SD  & $1^+_2$ & 3.85 & 0.50 & 30 & 3.61 & 19.52 \\
$^{45}$Sc 
 & spherical & $7/2^-_1$ & 0.00 & 0.17 & 0 & 3.51 &  \\
 & ND2 & $3/2^-_3$ & 2.18 & 0.45 & 35 & 3.62 &  \\
 & SD  & $3/2^+_2$ & 3.41 & 0.50 & 30 & 3.64 & \\
 \\
$^{48}_\Lambda$Sc 
 & spherical & $4^-_1$ & 0.00 & 0.15 & 8 & 3.53 & 20.12 \\
 & ND2 & $3^-_5$ & 5.98 & 0.45 & 59 & 3.63 & 19.69 \\
 & ND2 & $0^+_1$ & 3.96 & 0.35 & 15 & 3.59 & 19.88 \\
$^{47}$Sc 
 & spherical & $7/2^-_1$ & 0.00 & 0.15 & 8 & 3.54 &  \\
 & ND2 & $5/2^-_3$ & 5.55 & 0.45 & 59  & 3.65 &  \\
 & ND2 & $1/2^+_1$ & 3.72 & 0.40 & 17  & 3.63 & 
    \end{tabular}
  \end{ruledtabular}
\end{table*}

In odd Sc isotopes, it has been discussed, from the late 1960s, that the disappearance of the magic number $N=20$ and the coexistence of the different deformations in the ground-state energy region, $i.e.$ the appearance of the low-lying positive-parity states due to the promotion of a proton from the $sd$-shell \cite{PPS91.310(1967),NPA110.429(1968),PLB32.451(1970),PLB37.366(1971),Phys.Scr8.135(1973),NuovoCim.26A.25(1975),NPA262.317(1976),Act.Phys.Pol.B16.847(1985)}. 
Furthermore, it is expected that many-particle many-hole ($mp$-$mh$) states exist within the small excitation energy. 
Especially, Sc isotopes could have largely deformed states with a similar nucleon configuration to the SD states of $^{40}$Ca. 
In this Section, we focus on the response of the $\Lambda$ addition to $^{45}$Sc and $^{47}$Sc.

\subsection{$mp-mh$ and SD states}

In $^{45}$Sc and $^{47}$Sc, we calculate the energy curves as a function of $\beta$ for the negative- and positive-parity states. In Fig. \ref{fig:fig2.eps} (a)-(d), it is found that three (two) energy curves with the different minima are obtained for the negative- (positive-) parity state. 
As shown in Fig. \ref{fig:fig2.eps}(a), we obtain one spherical energy minimum and two ND minima with different deformations in the negative-parity states of $^{45}$Sc and $^{46}_\Lambda$Sc. In Fig. \ref{fig:fig2.eps}(b), we predict the ND and SD minima as the positive-parity states of $^{45}$Sc and $^{46}_\Lambda$Sc, while only ND minima are obtained in $^{47}$Sc and $^{48}_\Lambda$Sc.
To analyze the nucleon configuration of these energy curves, we calculate the numbers of the deformed harmonic oscillator quanta $N^p$ and $N^n$, defined by Eq. (\ref{NumHOQ}), using the wave functions on each energy curve. The $N^p$ ($N^n$) shows the $sd$-shell closed and/or particle-hole configurations of the protons (neutrons). 
In Fig. \ref{fig:fig3.eps}, it is found that the wave functions on each energy curve have the same nucleon configuration. Since the lowest value allowed by the Pauli principle is $N^p = 33$ and $ N^n = 42$  ($N^p = 33$ and $ N^n = 48$) for $^{45}$Sc ($^{47}$Sc), the spherical minimum of $^{45}$Sc ($^{47}$Sc) should have the $[(pf)^1]_\pi [(pf)^4] _\nu$ ($[(pf)^1]_\pi [(pf)^6] _\nu$) configuration. 
In Fig. \ref{fig:fig3.eps}, we see that the $N^p$ and/or $N^n$ values of the other curves, ND1, ND2 and SD, in $^{45}$Sc and $^{47}$Sc are different from the Pauli allowed values, which indicates that these energy curves have $mp$-$mh$ configurations. 
Among them, the energy curve denoted as SD in Fig. \ref{fig:fig2.eps}(b) is considered to have the $[(sd)^{-3} (pf)^4]_\pi [(sd)^{-2} (pf)^6] _\nu$ configuration, in which the proton configuration is similar to that of the SD states in $^{40}$Ca. 
Furthermore, to see deformation of ND1, ND2, and SD in $^{45}$Sc visually, we illustrate the density distributions of them in Fig. \ref{fig:dens.eps}. We see the large quadrupole deformation as $\beta$ increases. By the addition of a $\Lambda$ particle to these states, it is expected that the $\Lambda$-separation energy $B_\Lambda$ is different depending on the deformations as discussed for $^{41}_\Lambda$Ca.

Next, we calculate the energy curves of $^{46}_\Lambda$Sc and $^{48}_\Lambda$Sc, and the $\Lambda$ single particle energy $\epsilon_\Lambda$ as function of $\beta$.
The solid curves in Fig. \ref{fig:fig2.eps}(a)-(d) show that the $\Lambda$ hyperon generates the energy curves of $^{46}_\Lambda$Sc and $^{48}_\Lambda$Sc corresponding to the core nuclei. 
Figure \ref{fig:fig2.eps}(e)-(h) show the $\epsilon_\Lambda$ as a function of $\beta$ for the negative- and positive-parity states of $^{46}_\Lambda$Sc and $^{48}_\Lambda$Sc, and it is found that the behaviour of $\epsilon_\Lambda (\beta)$ is quite similar to that in $^{41}_\Lambda$Ca. 

To discuss the difference of the $\Lambda$-separation energy $B_\Lambda$, we perform the GCM calculation for $^{45}$Sc and $^{47}$Sc and the corresponding hypernuclei. 
Here we focus on the ground, SD and largely deformed ND states, which correspond to the spherical, SD and ND2 minima shown in Fig. \ref{fig:fig2.eps}(a)-(d), respectively.
Figure \ref{fig:fig4.eps} and Tab.\ref{Tab:table5} show the $B_\Lambda$ of the spherical and deformed states of $^{46}_\Lambda$Sc and $^{48}_\Lambda$Sc. 
In Fig. \ref{fig:fig4.eps} and Tab.\ref{Tab:table5}, it is found that the $B_\Lambda$ is the largest in the spherical states in $^{46}_\Lambda$Sc and $^{48}_\Lambda$Sc. And it is also found that the $B_\Lambda$ becomes smaller, as the nuclear quadrupole deformation $\beta$ is increased. This trend of $B_\Lambda$ is consistent with the behavior of $\epsilon_\Lambda (\beta)$ shown in Fig. \ref{fig:fig2.eps}(e)-(h), and is quite similar to the $^{41}_\Lambda$Ca shown in Fig. \ref{fig:fig1.eps}. 

Let us discuss the changes of matter radii and nuclear deformations by adding a $\Lambda$ hyperon.
In Tab.\ref{Tab:table5}, we see the almost no change of the matter r.m.s. radius and the nuclear quadrupole deformation $\beta$ of each state by adding a $\Lambda$ hyperon. 
Hence, a $\Lambda$ hyperon does not change the radius and deformations of nuclear part significantly in $^{46}_\Lambda$Sc and $^{48}_\Lambda$Sc.

\subsection{Excitation spectra}

\begin{figure*}
  \begin{center}
    \includegraphics[keepaspectratio=true,width=172mm]{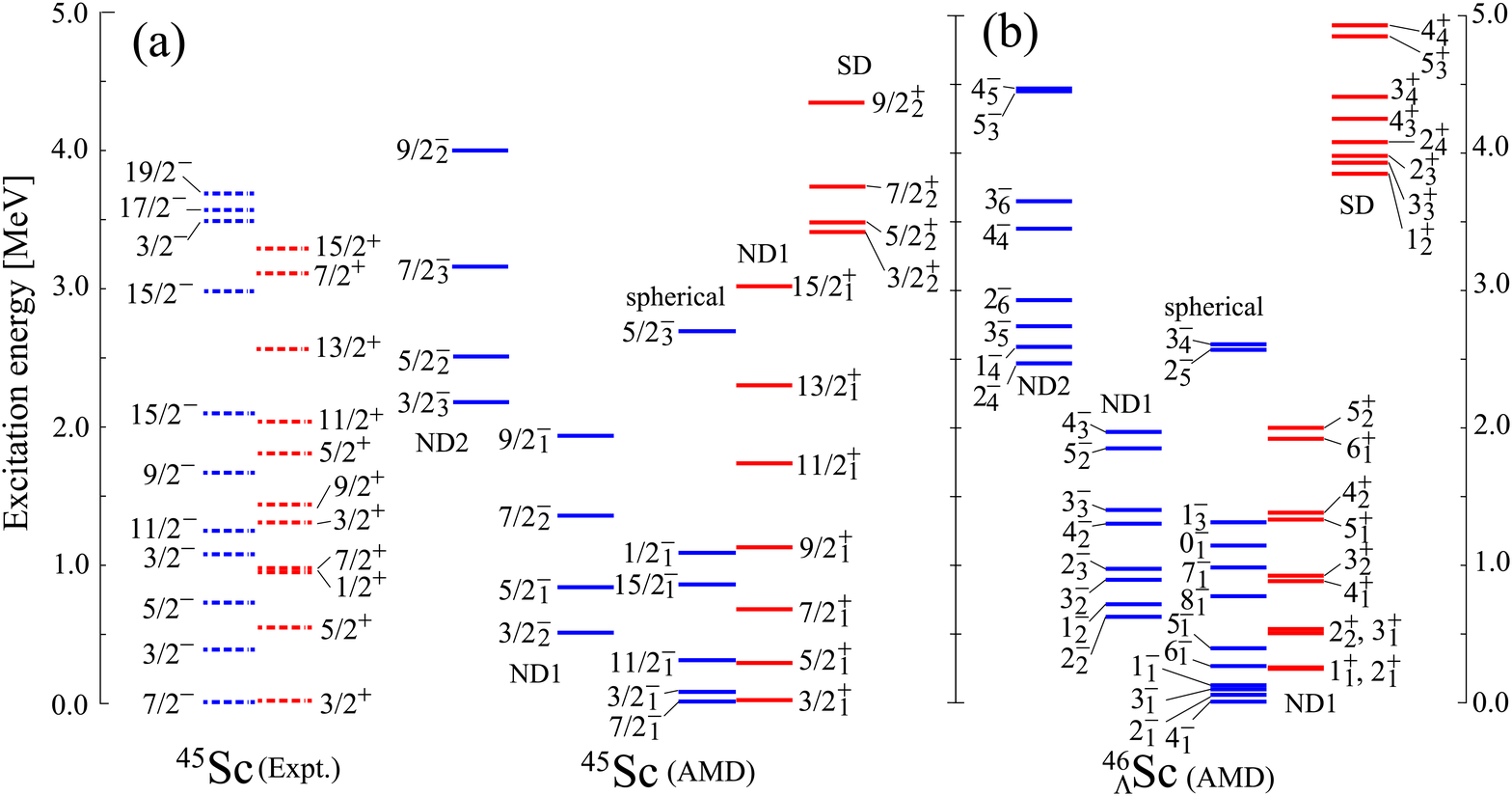}
  \end{center}
  \caption{(Color online) Calculated excitation spectra of $^{45}$Sc and $^{46}_\Lambda$Sc.}
  \label{fig:fig5.eps}
\end{figure*}

\begin{figure*}
  \begin{center}
    \includegraphics[keepaspectratio=true,width=172mm]{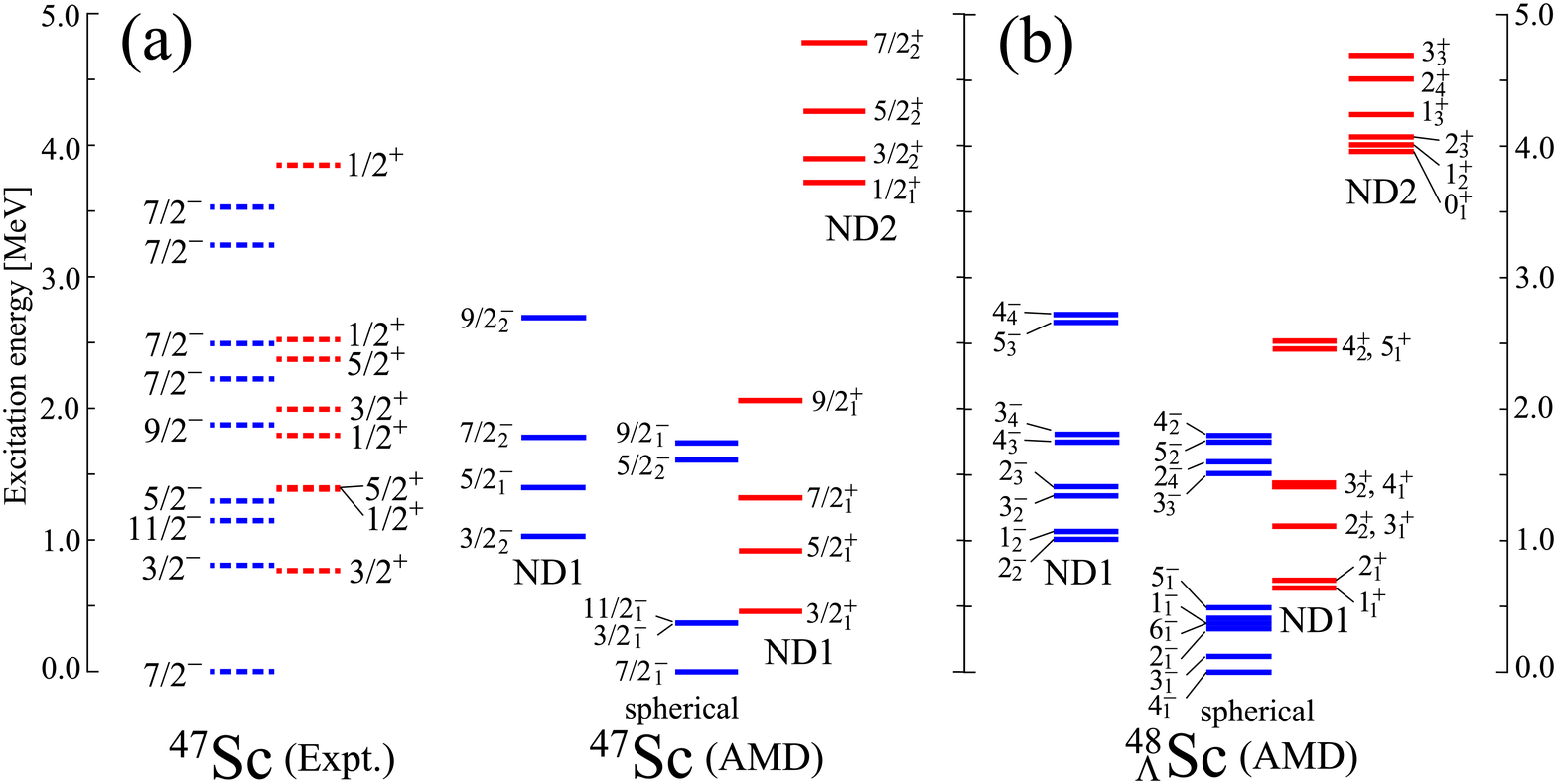}
  \end{center}
  \caption{(Color online) Same as Figure \ref{fig:fig5.eps} for $^{47}$Sc and of $^{48}_\Lambda$Sc.}
  \label{fig:fig6.eps}
\end{figure*}

\begin{table}
  \caption{Calculated $B(E2)$ [e$^2$ fm$^4$] values for $^{45}$Sc and $^{47}$Sc. Values in parentheses are observed data.}
  \label{Tab:table4}
  \begin{ruledtabular}
  \begin{tabular}{ccccccc}
  \multicolumn{3}{c}{$^{45}$Sc} & & \multicolumn{3}{c}{$^{47}$Sc} \\
$J_i$ & $J_f$ & $B(E2)$ &  & $J_i$ & $J_f$ & $B(E2)$ \\
\cline{1-3} \cline{5-7}
spherical &  &  &  & spherical &  &  \\
$11/2^-_1$ & $7/2^-_1$ & 56 (108 \cite{EPJA2.157(1998)}) &  & $11/2^-_1$ & $7/2^-_1$ & 37 \\
$15/2^-_1$ & $11/2^-_1$ & 33 ($<$818 \cite{EPJA2.157(1998)}) &  & $3/2^-_1$ & $7/2^-_1$ & 87 \\
$3/2^-_1$ & $7/2^-_1$ & 132 &  & $5/2^-_2$ & $3/2^-_1$ & 6 \\
$1/2^-_1$ & $3/2^-_1$ & 111 &  & $5/2^-_2$ & $7/2^-_1$ & 1 \\
$5/2^-_3$ & $1/2^-_1$ & 69 &  & $9/2^-_1$ & $5/2^-_2$ & 7 \\
$5/2^-_3$ & $3/2^-_1$ & 20 &  & $9/2^-_1$ & $7/2^-_1$ & 7 \\
ND1(-) &  &  &  & ND1(-) &  &  \\
$5/2^-_1$ & $3/2^-_2$ & 558 &  & $5/2^-_1$ & $3/2^-_2$ & 319 \\
$7/2^-_2$ & $3/2^-_2$ & 218 &  & $7/2^-_2$ & $3/2^-_2$ & 263 \\
$7/2^-_2$ & $5/2^-_1$ & 325 &  & $7/2^-_2$ & $5/2^-_1$ & 238 \\
$9/2^-_1$ & $5/2^-_1$ & 331 &  & $9/2^-_2$ & $5/2^-_1$ & 204 \\
$9/2^-_1$ & $7/2^-_2$ & 203 &  & $9/2^-_2$ & $7/2^-_2$ & 52 \\
ND2(-) &  &  &  & ND2(-) &  &  \\
$5/2^-_2$ & $3/2^-_3$ & 263 &  & $7/2^-_3$ & $5/2^-_3$ & 542 \\
$7/2^-_3$ & $3/2^-_3$ & 34 &  & $9/2^-_3$ & $5/2^-_3$ & 153 \\
$7/2^-_3$ & $5/2^-_2$ & 172 &  & $9/2^-_3$ & $7/2^-_3$ & 457 \\
$9/2^-_2$ & $5/2^-_2$ & 120 &  & $11/2^-_2$ & $7/2^-_3$ & 259 \\
$9/2^-_2$ & $7/2^-_3$ & 230 &  & $11/2^-_2$ & $9/2^-_3$ & 350 \\
$11/2^-_2$ & $7/2^-_3$ & 61 &  &  &  &  \\
$11/2^-_2$ & $9/2^-_2$ & 36 &  &  &  &  \\
$13/2^-_1$ & $9/2^-_2$ & 19 &  &  &  &  \\
$13/2^-_1$ & $11/2^-_2$ & 126 &  &  &  &  \\
 &  &  &  &  &  &  \\
ND1(+) &  &  &  & ND1(+) &  &  \\
$5/2^+_1$ & $3/2^+_1$ & 320 &  & $5/2^+_1$ & $3/2^+_1$ & 237 \\
 &  & (360$^{+332}_{-180}$\cite{NPA262.317(1976)}) &  &  &  & (432$^{+570}_{-293}$\cite{NPA262.317(1976)}) \\
$7/2^+_1$ & $3/2^+_1$ & 133 (62 \cite{EPJA2.157(1998)}) &  & $7/2^+_1$ & $3/2^+_1$ & 136 \\
$7/2^+_1$ & $5/2^+_1$ & 199 &  &  &  & (111$^{+81}_{-58}$\cite{NPA262.317(1976)}) \\
 &  & (115$^{+251}_{-91}$\cite{NPA262.317(1976)}) &  & $7/2^+_1$ & $5/2^+_1$ & 156 \\
$9/2^+_1$ & $5/2^+_1$ & 185 (179 \cite{EPJA2.157(1998)}) &  & $9/2^+_1$ & $5/2^+_1$ & 159 \\
$9/2^+_1$ & $7/2^+_1$ & 118 &  & $9/2^+_1$ & $7/2^+_1$ & 52 \\
 &  & (57$^{+65}_{-35}$\cite{NPA262.317(1976)}) &  & ND2(+) &  &  \\
$11/2^+_1$ & $7/2^+_1$ & 214  (216 \cite{EPJA2.157(1998)}) &  & $3/2^+_2$ & $1/2^+_1$ & 493 \\
$11/2^+_1$ & $9/2^+_1$ & 84 &  & $5/2^+_2$ & $1/2^+_1$ & 515 \\
$13/2^+_1$ & $9/2^+_1$ & 205 (151 \cite{EPJA2.157(1998)}) &  & $5/2^+_2$ & $3/2^+_2$ & 192 \\
$13/2^+_1$ & $11/2^+_1$ & 50 &  & $7/2^+_2$ & $3/2^+_2$ & 639 \\
$15/2^+_1$ & $11/2^+_1$ & 208 (223 \cite{EPJA2.157(1998)}) &  & $7/2^+_2$ & $5/2^+_2$ & 61 \\
$15/2^+_1$ & $13/2^+_1$ & 42 &  & $9/2^+_2$ & $5/2^+_2$ & 772 \\
SD(+) &  &  &  & $9/2^+_2$ & $7/2^+_2$ & 92 \\
$5/2^+_2$ & $3/2^+_2$ & 545 &  & $11/2^+_1$ & $7/2^+_2$ & 704 \\
$7/2^+_2$ & $3/2^+_2$ & 547 &  & $11/2^+_1$ & $9/2^+_2$ & 16 \\
$7/2^+_2$ & $5/2^+_2$ & 55 &  &  &  &  \\
$9/2^+_2$ & $5/2^+_2$ & 835 &  &  &  &  \\
$9/2^+_2$ & $7/2^+_2$ & 268 &  &  &  &  \\
$11/2^+_2$ & $7/2^+_2$ & 749 &  &  &  &  \\
    \end{tabular}
  \end{ruledtabular}
\end{table}

We discuss the excitation spectra of $^{45}$Sc and $^{47}$Sc and the corresponding hypernuclei, which are shown in Fig. \ref{fig:fig5.eps} and Fig. \ref{fig:fig6.eps}.
It is noted that all states shown in Fig. \ref{fig:fig5.eps} and Fig. \ref{fig:fig6.eps} are bound, because the lowest threshold energies of $^{45} $Sc and $^{47} $Sc are 6.9 MeV ($^{44}$Ca + $p$) and 8.5 MeV ($^{46}$Ca + $p$), respectively. 
Corresponding to $^{45}$Sc ($^{47}$Sc), the lowest threshold of $^{46}_\Lambda$Sc ($^{48}_\Lambda$Sc) is $^{45}_\Lambda$Ca + $p$ ($^{47}_\Lambda$Ca + $p$) and the threshold energy is expected to be higher than 6.9 MeV (8.5 MeV) in $^{46}_\Lambda$Sc ($^{48}_\Lambda$Sc). 
In $^{45}$Sc and $^{47}$Sc, by the GCM calculation, we obtain many excited states as shown in Fig. \ref{fig:fig5.eps}(a) and Fig. \ref{fig:fig6.eps}(a). 
By the GCM-overlap analysis, the negative- and positive-parity states of $^{45}$Sc and $^{47}$Sc are assigned as spherical, ND1, ND2 and SD states, depending on the major components of the wave function in the curves in Fig. \ref{fig:fig2.eps}(a)-(d). 
For example, in $^{45}$Sc, the states denoted as spherical in Fig. \ref{fig:fig5.eps}(a) are dominantly generated by the wave functions on the spherical minimum shown in Fig. \ref{fig:fig2.eps}(a). Similarly, the ND1 minimum shown in Fig. \ref{fig:fig2.eps}(a) dominantly contributes the positive-parity ND1 states in Fig. \ref{fig:fig5.eps}(a). 
Corresponding to the SD curve, the SD states appear in $^{45}$Sc. 
In Tab. \ref{Tab:table4}, it is found that the ND1, ND2 and SD states in each nucleus are connected by the large $B(E2)$ values, which is a clear sign of the existence of the rotational bands. In $^{45}$Sc, it is expected that the existence of the SD and ND bands can be confirmed by the observations of the inter- and intra-band transitions such as $B(E2)$ through the $^{24}$Mg + $^{24}$Mg fusion-evaporation reaction experiment \cite{Ideguchi}.  

Figure \ref{fig:fig5.eps}(b) shows the calculated excitation spectra of $^{46}_\Lambda$Sc. 
It shows that the ground and excited states of $^{46}_\Lambda$Sc with a $\Lambda$ hyperon are obtained corresponding to the spherical, ND1, ND2 and SD states of $^{45}$Sc.  
Focusing on the SD states of $^{46}_\Lambda$Sc, it is found from Fig. \ref{fig:fig5.eps} and Tab. \ref{Tab:table5} that the SD states are shifted up by about 440 keV. 
We see the similar shift up of the other deformed states. For example, the $3/2^-_3$ (ND2) state of $^{45}$Sc with $\beta = 0.45$ is shifted up in $^{46}_\Lambda$Sc by about 290 keV. These shifts reflect the difference of $B_\Lambda$ shown in the left panel of Fig. \ref{fig:fig4.eps}.

In $^{48}_\Lambda$Sc, we predict the existence of the ND1 and ND2 states with various deformations. 
It is found, from Fig. \ref{fig:fig6.eps}(b) and Tab. \ref{Tab:table5}, that the addition of a $\Lambda$ hyperon causes the similar modification of the excitation spectra as $^{46}_\Lambda$Sc, in which the $\Lambda$ hyperon shifts the largely deformed states up. For example, the Tab. \ref{Tab:table5} shows that the excitation energy of the $5/2^-_3$ with $\beta = 0.45$ is shifted up by 430 keV in $^{48}_\Lambda$Sc. In this way, we find that the shifted energy is larger as $\beta$ increases.

\subsection{Response to the $7/2^-$ and $3/2^+$ states by addition of a $\Lambda$ particle}

\begin{table}
  \caption{The electric quadrupole moment ($Q$ in unit of e fm$^2$) of the $7/2^-$ and $3/2^+$ states in $^{45}$Sc and $^{47}$Sc. The values in parentheses are the observed data.}
  \label{Tab:table3}
  \begin{ruledtabular}
  \begin{tabular}{ccccccccc}
  \multicolumn{4}{c}{$^{45}$Sc} & & \multicolumn{4}{c}{$^{47}$Sc}\\
   & $J^\pi$ & $Q$ & && & $J^\pi$ & $Q$ & \\
  \cline{1-4} \cline{5-9}
  & $7/2^-$ & -15 (-22.0$^{-0.2}_{-0.2}$ \cite{DataTable90.75(2005)})  & && & $7/2^-$ & -13  (-22$^{+3}_{-3}$ \cite{DataTable90.75(2005)}) &  \\
  & $3/2^+$ & +18 (+28$^{+5}_{-5}$ \cite{JPGNPP38.025104(2011)}) & && & $3/2^+$ & +16 &  \\
  \end{tabular}
  \end{ruledtabular}
\end{table}

\begin{figure*}
  \begin{center}
    \includegraphics[keepaspectratio=true,width=172mm]{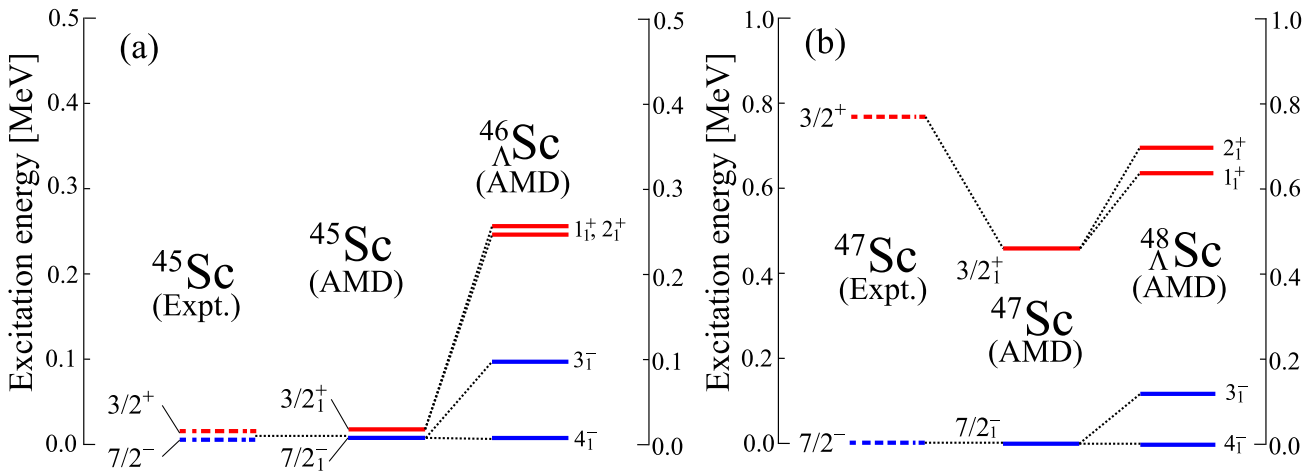}
  \end{center}
  \caption{(Color online) (a) The ground ($7/2^-_1$) and the lowest positive-parity states ($3/2^+_1$) of $^{45}$Sc, and the corresponding states in $^{46}_\Lambda$Sc. (b) Same as (a) for $^{47}$Sc and $^{48}_\Lambda$Sc.}
  \label{fig:fig7.eps}
\end{figure*}

Here, we focus on the ground and low-lying states $7/2^-_1$ and $3/2^+_1$ of $^{45}$Sc.
It is well known experimentally that the $3/2^+_1$ state lies at slightly higher than the $7/2^-_1$ state and the energy difference between them is 12 keV. 
This is inconsistent with the shell-model picture, in which a $3/2^+$ state should be located by a few MeV with respect to the ground state $7/2^-_1$. 
This inconsistency has been extensively discussed theoretically. Then, we understand in terms of deformation by the Nilsson model as follows \cite{Phys.Scr8.135(1973),NuovoCim.26A.25(1975),NPA262.317(1976),Act.Phys.Pol.B16.847(1985)}: as increasing deformation, the Nilsson $K^\pi = 3/2^+$ state from the $d_{3/2}$ becomes close to the $K^\pi = 1/2^-$ state from the $0f_{7/2}$. As a result, the $3/2^+$ state becomes lower and very close to the ground state $7/2^-$. 
We found that deformation in the $3/2^+_1$ state is larger than that in the $7/2^-_1$ state. The calculated quadrupole deformations of $7/2^-_1$ and $3/2^+_1$ are $\beta = 0.17$ and $\beta = 0.27$, respectively.
The difference of the deformations between them is also seen in the $Q$ momentum and $B(E2)$ strength. 
The calculated $Q$ momenta of the $7/2^-_1$ and $3/2^+_1$ states are listed in Tab. \ref{Tab:table3}, which are not inconsistent with the observed data within the error bar.

It is expected that the responses to $\Lambda$ participation in the $7/2^-_1$ and $3/2^+_1$ states are different due to the difference in their deformations. 
In $^{46}_\Lambda$Sc, calculated $B_\Lambda$ in the $1^+_1$ and $2^+_1$ states is 19.71 MeV, while $B_\Lambda$ in the $3^-_1$ and $4^-_1$ states is 19.95 MeV, since the $\Lambda$-separation energy $B_\Lambda$ in more deformed states is smaller than that in less deformed state. As a result, as shown in Fig. \ref{fig:fig7.eps}(a), the $1^+_1$ and $2^+_1$ states are shifted up than the $3^-_1$ and $4^-_1$ states and energy difference is much larger by $\sim 0.2$ MeV. 
The same behavior is seen in $^{47}$Sc and $^{48}_\Lambda$Sc (see in Fig. \ref{fig:fig7.eps} (b)), and the increase of the excitation energies is also around 0.2 MeV. 
In Fig. \ref{fig:fig7.eps}, we see energy splitting for the ground state doublet, $3^-_1 - 4^-_1$, and for the excited doublet, $1^+_1 -2^+_1$, are quite different with each other. The reason is as follows: 
The number of proton in the highest orbit ($pf$-shell) in $7/2^-_1$ state is one, while that in the $3/2^+_1$ state is two. The energy splitting in the $7/2^-_1$ state comes from spin-spin term in $\Lambda N$ interaction between one proton and $\Lambda$ in $0s$-orbit. If we use even-state spin-spin term for $3^-_1 - 4^-_1$ state, the ground states of $^{46}_\Lambda$Sc and $^{48}_\Lambda$Sc is $3^-$. When we use odd-state spin-spin term, due to the attraction of the triplet-state than the singlet-state, $4^-$ becomes the ground state. As a result, the energy splitting for the $4^-_1 - 3^-_1$ doublet is about 90 keV. On the other hand, spin of two protons in the $3/2^+_1$ state is 0 (spin anti-parallel). Then, spin-spin part is almost cancelled out and energy splitting for the $1^+_1 - 2^+_1$ double state is negligibly small. 

If we could observe these energy difference experimentally, it would be helpful to obtain information on $\Lambda N$ spin-dependent part. However, since these two energy splittings are less than 100 keV, then it is might be difficult to be observed experimentally.

\section{Summary} 

We have studied the deformations of $^{40}$Ca, $^{45}$Sc and $^{47}$Sc, and those of the corresponding hypernuclei. 
For $^{45}$Sc and $^{47}$Sc, the deformed-basis AMD + GCM was adopted; the same model was applied to investigate the ND and SD bands of $^{40}$Ca in Ref. \cite{PRC76.044317(2007)}. We also investigated the ND and SD states of $^{41}_\Lambda$Ca, $^{46}_\Lambda$Sc and $^{48}_\Lambda$Sc using the HyperAMD. As the effective interactions, the Gogny D1S was adopted for $NN$ sector, and the latest version of the ESC08c $\Lambda N$ potential was employed. 
Major points to be emphasized are as follows:

(i) In $^{40}$Ca, we obtained three $0^+$ states corresponding to the observed spherical ground, ND and SD states, which are identical to those reported in Ref. \cite{PRC76.044317(2007)}. By addition of a $\Lambda$ particle to these states, it was found that the calculated $\Lambda$-separation energy $B_\Lambda$ was depending on the degree of deformations: $B_\Lambda = 19.45$ MeV for the ground state ($1/2^+_1$), $B_\Lambda = 19.15$ MeV for the ND state ($1/2^+_2$) and $B_\Lambda = 18.01$ MeV for the SD state ($1/2^+_3$) in $^{41}_\Lambda$Ca. 

(ii) In the core nuclei $^{45}$Sc and $^{47}$Sc, we obtained many excited states with different ($\beta$,$\gamma$) deformations, which have $mp$-$mh$ configurations. Among these states, we predicted, for the first time, that $^{45}$Sc had SD states in which protons had the same configuration as the SD states of $^{40}$Ca. By addition of a $\Lambda$ particle, the resultant hypernuclei, $^{46}_\Lambda$Sc and $^{48}_\Lambda$Sc, had $mp$-$mh$ states with a $\Lambda$ in $s$-orbit. Furthermore, we found that the calculated $B_\Lambda$ also depended on the degree of deformation: for example, in $^{46}_\Lambda$Sc, $B_\Lambda$ in the SD state was 19.52 MeV, while $B_\Lambda = 19.95$ MeV in the ground state. 

(iii) We focused on the level structure of $7/2^-_1$ and $3/2^+_1$ states in $^{45}$Sc and $^{47}$Sc; these two states were close to each other. Since the deformation $\beta$ of each state was different, the calculated $B_\Lambda$ was different from each other. 
As a result, the corresponding states ($1^+_1$ and $2^+_1$) in $^{46}_\Lambda$Sc and $^{48}_\Lambda$Sc were shifted up than the $3^-_1$ and $4^-_1$ states. 

To see the above interesting phenomena in $^{46}_\Lambda$Sc and $^{48}_\Lambda$Sc, we hope to perform high resolution experiments, $^{46}$Ti$(e,e'K^+)$$^{46}_\Lambda$Sc and $^{48}$Ti$(e,e'K^+)$$^{48}_\Lambda$Sc at JLab in the future.

\begin{acknowledgments}
The authors would like to thank Prof. S. N. Nakamura, Dr. Y. Fujii, Dr. K. Tsukada and Mr. T. Gogami for valuable discussions and information on the experimental project at JLab. They are also thankful to Professors E. Ideguchi, T. Motoba and Prof. S.-G. Zhou for helpful discussions and encouragement. This work is supported in part by the Grants-in-Aid for Scientific Research on Innovative Areas from MEXT (No.2404: 24105008). M. I. is supported by the Special Postdoctoral Researcher Program of RIKEN. The numerical calculations were performed on the HITACHI SR16000 at KEK.
\end{acknowledgments}

\appendix*
\section{Parameters of the YNG-ESC08c interaction}

As for YN interaction employed in this paper, we use the YNG interaction derived from ESC08c. The parameters $a_i$, $b_i$ and $c_i$, and $\beta_i$ in Eq. (\ref{YNG}) are summarized in Table \ref{Tab:app1} and Table \ref{Tab:app2} for the central, and SLS and ALS parts of the $G$-matrix interaction, respectively.

\begin{table}
  \caption{Parameters of $\Lambda N$ $G$-matrix interaction represented by the three range Gaussian forms given in Eq. (\ref{YNG}) in case of ESC08c. $a_i$, $b_i$ and $c_i$ are given in MeV, and $\beta_i$ is given in fm for each $i$.}
  \label{Tab:app1}
  \begin{ruledtabular}
  \begin{tabular}{ccccc}
  &  $\beta_i$ & 0.5  &  0.9   &  2.0   \\
  \hline
        & $a_i$  &  --3892.0  &    471.5  & --1.756  \\
  $^1E$ & $b_i$  &    7401.0  &  --1132.0  &   0.0    \\
        & $c_i$  &  --2821.0  &    446.4  &   0.0    \\
  \\
        & $a_i$  &  --2135.0  &    229.7  & --1.339  \\
  $^3E$ & $b_i$  &    4705.0  &  --727.1  &   0.0    \\
        & $c_i$  &  --1939.0  &    319.8  &   0.0    \\
  \\
        & $a_i$  &    1785.0  &    103.6  & --0.7956  \\
  $^1O$ & $b_i$  &    9.656  &    14.47  &   0.0    \\
        & $c_i$  &  --16.57  &    14.57  &   0.0    \\
  \\
        & $a_i$  &    1838.0  &  --13.82  & --1.001  \\
  $^3O$ & $b_i$  &  --1966.0  &  --233.9  &   0.0    \\
        & $c_i$  &    682.7  &    192.8  &   0.0    \\
  \end{tabular}
  \end{ruledtabular}
\end{table}

\begin{table}
  \caption{Same as Table \ref{Tab:app1} for the SLS and ALS parts.}
  \label{Tab:app2}
  \begin{ruledtabular}
  \begin{tabular}{ccccc}
  & $\beta_i$ & 0.4  &  0.8 &  1.2  \\
  \hline
      & $a_i$ & --11470.0 &   299.3 & --2.386  \\
  SLS & $b_i$ &   21880.0 & --744.1 &   0.0    \\
      & $c_i$ & --8962.0  &   294.4 &   0.0    \\
  \\
      & $a_i$ &   2084.0 &   9.536 &  1.851 \\
  ALS & $b_i$ & --1549.0 &   32.55 &  0.0   \\
      & $c_i$ &   578.1 & --14.52 &  0.0   \\
  \end{tabular}
  \end{ruledtabular}
\end{table}


\newpage 
\bibliography{Sc-hyper} 

\end{document}